\documentclass[prd,aps,twocolumn,superscriptaddress,amsmath,amssymb]{revtex4-1}

\usepackage[colorlinks=true,citecolor=blue,filecolor=blue,linkcolor=blue,urlcolor=blue]{hyperref}

\usepackage{graphicx}           
\usepackage{array}              
\usepackage{placeins}           
\usepackage[dvipsnames]{xcolor}  
\usepackage{tikz}               
\usetikzlibrary{shapes, decorations.pathreplacing}  

\usepackage{amsmath}            
\usepackage{amssymb}            
\usepackage{siunitx}            
\usepackage{xspace}             
\usepackage{bm}                 
\usepackage{ifthen}

\usepackage[T1]{fontenc}
\usepackage[utf8]{inputenc}
\usepackage[english]{babel}
\usepackage{csquotes}           

\usepackage{lipsum}             
\usepackage{verbatim}           
\usepackage{pgf}                
\usepackage{xcolor}             
\usepackage[utf8]{inputenc}     
\usepackage{microtype}          

\newcommand{\stockholm}{\affiliation{Oskar Klein Centre, Department of Physics, Stockholm University, AlbaNova, Stockholm SE-10691, Sweden}}

\usepackage{mathtools}
\DeclarePairedDelimiter\ceil{\lceil}{\rceil}
\DeclarePairedDelimiter\floor{\lfloor}{\rfloor}

\newcommand\definable[1]{\textcolor{ForestGreen}{#1}}

\newcommand{\tsone}{\text{S1}}
\newcommand{\tstwo}{\text{S2}}
\newcommand{\phot}{\text{ph}}
\newcommand{\phe}{\text{pe}}
\newcommand{\elec}{\text{el}}
\newcommand{\dete}{\text{det}}
\newcommand{\produ}{\text{prod}}
\newcommand{\fd}{\textsc{flamedisx}}
\newcommand{\roughly}{\sim\!}
\newcommand{\energysymbol}{E}

\newcommand{\maxsigma}{\ensuremath{\sigma_\text{max}}}

\begin{document}

\title{Finding Dark Matter Faster with Explicit Profile Likelihoods}
\author{J.~Aalbers}\email[]{jelle.aalbers@fysik.su.se}\stockholm
\author{B.~Pelssers}\email[]{bart.pelssers@fysik.su.se}\stockholm
\author{V.C.~Antochi}\email[]{cristian.antochi@fysik.su.se}\stockholm
\author{P.L.~Tan}\email[]{pueh-leng.tan@fysik.su.se}\stockholm
\author{J.~Conrad}\email[]{conrad@fysik.su.se}\stockholm
\date{\today}

\begin{abstract}
Liquid xenon time-projection chambers are the world's most sensitive detectors for a wide range of dark matter candidates. We show that the statistical analysis of their data can be improved by replacing detector response Monte Carlo simulations with an equivalent deterministic calculation.
This allows the use of high-dimensional undiscretized models, yielding up to $\roughly 2$ times better discrimination of the dominant backgrounds. In turn, this could significantly extend the physics reach of upcoming experiments such as XENONnT and LZ, and bring forward a potential $5 \sigma$ dark matter discovery by over a year.
\end{abstract}


\maketitle

\section{Introduction}

\subsection{Motivation}
Astrophysical and cosmological measurements have established that about four-fifths of the mass of the universe's matter consists of dark matter \cite{bertone_dm, planck_2018}. Several experiments attempt to detect dark matter particles on earth \cite{undrauch_dd, rosz_dd, schumann_dd_2019}. Liquid xenon time projection chambers (LXe TPCs) lead the field of direct detection for many dark matter models \cite{sr1_prl, sd_prl, s2only_prl}, including the prominent category of models with dark matter mass around $\roughly \SI{100}{GeV/c^2}$ known as weakly interacting massive particles (WIMPs).

Along with other rare-event searches \cite{atlashiggs, cmshiggs, dayabay}, LXe TPCs use (profile) likelihood ratio tests to derive their final physics results, since these tests have (nearly) optimal statistical power \cite{neyman_pearson}. In this work, we suggest an improvement to how upcoming LXe TPC experiments such as 
XENONnT \cite{nt},
LZ \cite{lz_tdr},
and PandaX-4T \cite{px4t}
calculate their likelihood. We show this leads to a higher sensitivity through improved signal/background discrimination, and more robust results through supporting more simultaneous correlated nuisance parameters.

\subsection{Background}

Particle physics experiments often specify their detector response model implicitly through a Monte Carlo (MC) simulation. 
However, their inference uses (profile) likelihoods, which need the differential expected event rate -- the number of events expected in an infinitesimal volume around the event, in the space of observables such as signal amplitude or reconstructed position.
To obtain differential rates from a simulation, experiments use density estimation: usually, they histogram the \mbox{results} of a large simulation run, e.g.~$\mathcal{O}(10^{7})$ events. Such histogram-derived densities are known as \emph{templates}.

Parameters of the detector response model are implemented as configuration options of the simulation. During inference, an optimizer
explores many parameter-value combinations. To avoid re-doing simulations at every point during inference, experiments often precompute templates at strategically chosen points in the parameter space, and interpolate between them; linearly, or using more advanced techniques -- this is known as \emph{template morphing} \cite{morphing}.

To maximize sensitivity, an experiment would like to use all relevant simultaneous observables measured in an event. In particular, unless the detector response is homogeneous and constant in time, reconstructed position and time are relevant. This means templates must be high-dimensional histograms, which require a large number of simulations.

To derive robust results, experiments account for uncertainties as `nuisance' (i.e.~additional) parameters of the likelihood, which are profiled during inference (or marginalized, in Bayesian methods).
The more nuisance parameters are used, the more templates must be precomputed. In particular, correlated parameters require templates precomputed on a multidimensional grid, causing an exponential scaling of cost with number of parameters. 

These requirements have an escalating price in terms of computation. Experiments therefore often make compromises that decrease their physics potential, such as discretizing the model in a few spatial- or temporal `bins', or reduce robustness, such as considering only a few key or aggregate uncertainties as nuisance parameters.

\subsection{Concept and outline}
\label{section:outline}

In this paper, we show how to compute the differential expected event rate for events in LXe TPCs directly, without MC simulations or templates. The result is \emph{not} an approximation of an MC. Rather, we directly obtain the result of running an MC simulation with infinite statistics to fill an infinitesimally binned histogram, separately for the detector conditions appropriate for each individual event.

We build on the basic idea that every MC simulation approximates an integral or sum. For example, take a trivial simulation such as:
\begin{align}\begin{split}
    N &\sim \mathrm{Poisson}(\lambda) \\
    S &\sim \mathrm{Gauss}(\mu = N, \sigma = 0.1 \sqrt{N}) .
\end{split}\end{align}
This might represent an experiment observing signals of size $S$, obtained by Gaussian smearing from a number of photons $N$, which is in turn is sampled from a Poisson distribution with mean $\lambda$. 
To compute the probability density $P(s)$ at one or more observations, we could run the simulation many times, populate a histogram, and look up the estimated density at each observed $s$. Alternatively, we could compute the following sum:
\begin{align}\begin{split}
    P(s) &= \sum_n P(s|n) P(n) \\
    &= \sum_n \mathrm{Gauss}(s-n, 0.1 \sqrt{n}) \mathrm{Poisson}(n | \lambda) , 
\end{split}\end{align}
where $\mathrm{Poisson}(n | \lambda) = \lambda^n e^{-\lambda}/{n!}$ and $\mathrm{Gauss}(x, \sigma) = \exp{(-x^2/(2\sigma^2)}/\sqrt{2 \pi \sigma^2}$. This method works even if the mean of the Gaussian distribution is a complicated function of the number of detected photons, e.g. involving an externally determined response map.
The sum runs over all non-negative integers, but evaluating only e.g.~$n \in \{\floor{s-0.5\sqrt{s}}, \floor{s-0.5\sqrt{s}} + 1, ..., \ceil{s+0.5\sqrt{s} \,}\}$ will give accurate results except for $\gtrsim 5 \sigma$ outliers.

For the LXe emission and detector response model, the summation is more complicated to construct, and sensible bounds are more difficult to estimate -- but the principle is the same. The summation becomes a large matrix/tensor multiplication, implemented using TensorFlow \cite{tensorflow}, to take advantage of GPU acceleration and automatic differentiation. The latter gives access to the gradient and Hessian matrix in the space of nuisance parameters, which makes inference with a high number of nuisance parameters practical.

We present an open-source Python package \fd~\cite{fd_github} which integrates the direct differential rate computation in an inference framework. Instead of specializing to one experiment's choice of LXe model, we provide a framework in which functions such as the parametrizations of yields can be easily customized. 
Our model skeleton is inspired by, though not exactly equivalent to, NEST \cite{nest_paper, nest_software} and models used in XENON1T \cite{sr0_prl, analysis_paper_1, analysis_paper_2, aalbers_phd}.
\fd~includes utilities for statistical inference with the MINUIT and SciPy optimizers \cite{minuit, iminuit, scipy_paper}.
Finally, we show that \fd's high-dimensional likelihood gives up to a factor $\roughly 2$ better discrimination of the dominant electronic recoil background when compared with a classical 2d MC-template approach, and a corresponding increase in physics reach for dark matter searches.

Section \ref{section:tpcs} reviews the principle of LXe TPCs and their likelihoods as they are currently used. Section \ref{section:method} discusses \fd~'s method and models in detail. Section  \ref{section:results} discusses \fd's performance on a test case that aims to resemble future detectors such as XENONnT and LZ. We end with a short summary and outlook in section \ref{section:outlook}.

\section{LXe TPCs and their likelihoods}
\label{section:tpcs}

In this section, we briefly recapitulate the operational principle of LXe TPCs, and introduce notation used later in the paper.
A thorough exposition of LXe TPCs can be found in \cite{elena_tome, xenon1t_instrument, lz_instrument}.

\subsection{LXe TPCs}

\begin{figure}
    \centering
    \includegraphics[width=\columnwidth]{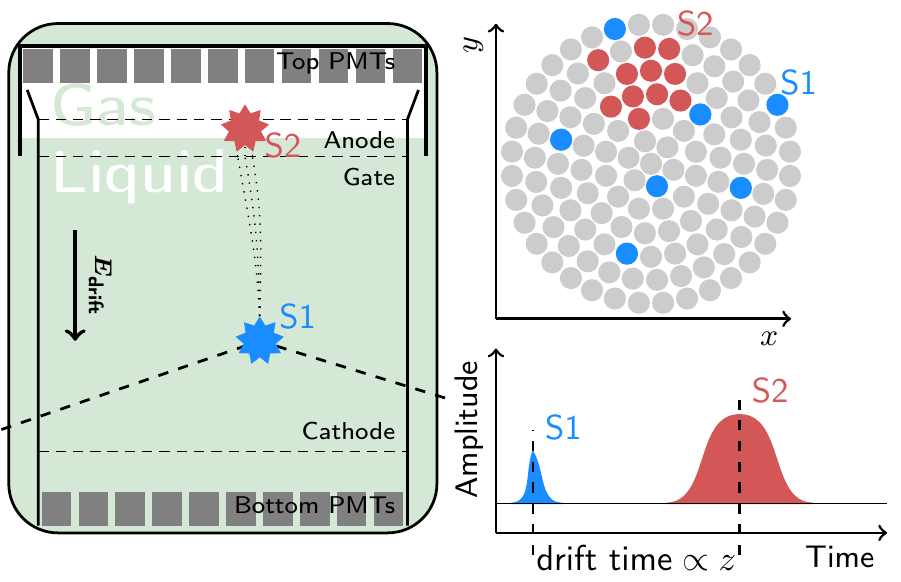}
    \caption{Working principle of LXe TPCs. Left panel: Sketch of a longitudinal cross-section of a LXe TPC, with the S1 and S2 interaction sites, PMT locations, and main electrodes indicated. Top right: Example hit pattern on the top array of an S1 (blue) and S2 (red). Bottom right: Sketch of a LXe TPC signal waveform (sum of the PMT signals). The S2 hit pattern is used to reconstruct the transverse $(x, y)$ position, and the drift time between S1 and S2 the event depth $z$.}
    \label{fig:tpc_diagram}
\end{figure}

TPCs used in modern dark matter searches are meter-sized cylinders filled with liquid xenon, with a few centimeter thick layer of gaseous xenon at the top. Figure \ref{fig:tpc_diagram} illustrates the working principle of such a LXe TPC.
Particles scatter off either the electrons or the nucleus of a xenon atom, resulting in electronic (ER) or nuclear recoils (NR), respectively. 
Recoils create excited xenon dimers, which decay within $\mathcal{O}(\SI{10}{ns})$, emitting UV light observed as a signal (S1) by the two arrays of photomultipliers (PMTs) at the top and bottom of the detector. Electron/ion pairs are also created in the recoil. The electrons are drifted up towards the gas by a $\mathcal{O}(\SI{100}{V/cm})$ electric field. The electrons then produce a larger secondary signal (S2) in the gaseous xenon under the influence of a $\mathcal{O}(\SI{10}{kV/cm})$ electric field. The light pattern of the S2 signal on the top PMT array indicates the horizontal $(x, y)$ position of the event; the drift time of the electrons (measured by the time delay between S1 and S2) indicates the depth $(z)$ of the interaction. The relative size of the S1 and S2 signals depends on the recoil's energy and type (ER or NR). This allows LXe TPCs to discriminate the NR signals expected for dark matter (and e.g. radiogenic neutron backgrounds) from ERs, primarily from radioactive contaminants in the LXe.

\subsection{Observables and likelihoods}

Likelihood tests require two key inputs: reconstructed quantities of observed events, and expected differential rates for signal and background sources.

LXe TPC likelihoods use a vector of observed quantities, $\bm{s}$, for each event that passes the selection criteria. The $\bm{s}$ usually includes the amplitudes S1 and S2 of the main detector signals in photoelectrons (PE), sometimes the reconstructed event position -- $x, y, z$ or $r, z, \phi$ depending on whether Cartesian or cylindrical coordinates are used -- and sometimes the measured absolute event time $t$. The likelihood is always expressed in directly observable quantities, both in this work and all implementations we are aware of in the field. For instance, the event positions are reconstructed positions, not presumed true positions.

Likelihoods also require the differential event rate $R^j(\bm{s})$ for different signal or background sources $j$. For example, if $\bm{s} = [\mathrm{S1}, \mathrm{S2}, x, y, z, t]$, $R^{j}$ relates to $\mu^{j}$, the total expected number of events after selections of the source $j$, as
\begin{equation}
    \label{eq:mu_definition}
    \mu^{j} = \rho \int d\mathrm{S1} \, d\mathrm{S2} \, dx \, dy \, dz \, dt 
    \; R^{j}(\mathrm{S1}, \mathrm{S2}, x, y, z, t)
    .
\end{equation}
Here $\rho$ is the liquid xenon density in the detector, so that $R^j$ has units of $\text{events}/(\mathrm{tonne} \times \mathrm{year} \times \mathrm{PE}^2)$. During inference, $\mu^j$ and $R^j$ depend on several parameters, such as the dark matter cross-section, the rates of different backgrounds, parameters of the LXe charge and light yield response functions, or experimental parameters such as electron lifetime.

To compare models against the observed data, LXe TPCs commonly use an extended unbinned likelihood:
\begin{equation}
    \label{eq:generic_l}
    L = 
    \mathrm{Poisson}(N_{\mathrm{tot}} | \mu) 
    \prod\limits_{i}^{\mathrm{events}} \; 
    \sum\limits_{j}^{\mathrm{sources}} 
    \frac{R^{j}(\bm{s}_{i})}{\mu}
    .
\end{equation}
Here, $\mathrm{Poisson}$ is the Poisson probability mass function, $\mu = \sum_j \mu^j$ is the total expected number of events from all sources, and $\mathrm{N}_\mathrm{tot}$ is the actual observed number of events. Eq.~\ref{eq:mu_definition} ensures that the sum calculates the probability density function of events in the total model. In practice, the logarithm of eq.~\ref{eq:generic_l} is used:
\begin{equation}
    \label{eq:generic_logl}
    \log L = - \mu + \sum\limits_{i}^{\mathrm{events}} \log \Bigg( \sum\limits_{j}^{\mathrm{sources}} R^{j}(\bm{s}_i) \Bigg)  + \mathrm{constant} .
\end{equation}
The constant depends on the dataset but not the model, and cancels in likelihood ratio inference. It is thus omitted.

The likelihood can have many parameters, but one type is worth highlighting: \emph{rate multipliers} $r^j$, which are unitless scales of both $\mu^j$ and $R^j$. That is, 
\begin{align}\begin{split}
    \label{eq:rate_multipliers}
    \mu^j(\theta) &= r^j \times m^j(\theta) \\
    R^j(\theta) &= r^j \times M^j(\theta),
\end{split}\end{align}
where $m^j(\theta)$ and $M^j(\theta)$ are the expected number of events and differential rate, respectively, for the source $j$ if the multiplying parameter $r^j = 1$, and $\theta$ encapsulates the other (nuisance) parameters in the likelihood, known as \emph{shape parameters}. The $r^j = 1$ represents a reference level, e.g. a cross-section of \SI{1e-47}{cm^2} for a WIMP, or a particular nominal level for the internal ER background. 

\subsection{Inhomogeneity and corrections}

LXe TPCs do not respond in the same way to signals produced at different positions in the detector or at different times during a science run. The S1 photon detection efficiency is generally highest close to the bottom PMT array, and lowest at the top edge of the liquid volume near the liquid-gas interface (which reflects photons impinging on it at a sufficiently large angle). The survival probability of S2 electrons decreases with depth because impurities in the LXe absorb drifting electrons, as quantified by the mean electron lifetime, $\tau$. The electron lifetime during a science run is often time-dependent. There are also other, usually smaller, inhomogeneities in the response of LXe TPCs, such as variations in the drift and extraction fields, PMT gains, PMT quantum efficiencies, etc.

\fd~operates on the full $(\mathrm{S1}, \mathrm{S2}, x, y, z, t)$ space of observables, and can thus take full account of these effects. The field currently uses template-based methods, for which, as mentioned above, using six dimensions is impractical, except by coarsely discretizing some dimensions. Instead, collaborations define space-and time dependent corrections for their signals, so they can use lower-dimensional likelihoods.
The corrected signals are called $(\mathrm{cS1}, \mathrm{cS2})$, or sometimes $(\mathrm{S1}_c, \mathrm{S2}_c)$. 
Using a template in $(\mathrm{cS1}, \mathrm{cS2})$, rather than $(\mathrm{S1}, \mathrm{S2})$, gives a much better signal/background discrimination.

Let $G_1(x, y, z, t)$ and $G_2(x, y, z, t)$ denote the mean expected signal in PE per released photon or electron, respectively, at a given position, $\mathrm{S1}/G_1$ and $\mathrm{S2}/G_2$ are asymptotically unbiased estimates of the number of photons and electrons released at the interaction site. These are usually multiplied by constant scale factors $g_1$ and $g_2$ to obtain the corrected signals:
\begin{align}\begin{split}
\mathrm{cS1}(\mathrm{S1}, x, y, z, t) &= \mathrm{S1} \frac{g_1}{G_1(x, y, z, t)} \\
\mathrm{cS2}(\mathrm{S2}, x, y, z, t) &= \mathrm{S2} \frac{g_2}{G_2(x, y, z, t)} .
\end{split}\end{align}
The $g_1$ and $g_2$ factors are chosen so that, for a homogeneously distributed source, $\mathrm{cS1} = \mathrm{S1}$ on average inside the fiducial volume, and $\mathrm{cS2} = \mathrm{S2}$ at the liquid-gas interface. To be precise:
\begin{align}\begin{split}
g_1 &= \frac{1}{V \delta T} \int dx \, dy \, dz \, dt \; G_1(x, y, z, t) \\
g_2 &= \frac{1}{A \delta T} \int dx \, dy \, dt \; G_2(x, y, z = 0, t) ,
\end{split}\end{align}
with the integrals running over the bounds of the fiducial volume $V$ and its projected (x, y) area $A$ respectively, and $\delta T$ the duration of the search.

\fd~uses the full space of observables, $(\mathrm{S1}, \mathrm{S2}, x, y, z, t)$, rather than just $(\mathrm{cS1}, \mathrm{cS2})$. Since $\mathrm{cS1}$ and $\mathrm{cS2}$ are direct functions of these, \fd~can still use efficiencies or other functions parametrized in these corrected variables.

Using the full observable space is preferable for two reasons. First, signal and background intensities and spectra often vary with $(x, y, z, t)$, even if the detector response was homogeneous, and even in the fiducial volume of the detector. Second, signal corrections only compensate for a change in the mean response, not for differences in resolution. For example, the electron detection efficiency decreases with depth in a TPC, worsening the S2 resolution further down the detector. For this reason, top regions of TPCs often show the best signal/background discrimination. If the analysis does not distinguish these from the bottom regions, it loses physics potential.

\section{Method}
\label{section:method}

\subsection{Overview}

\begin{figure}
    \centering
    \includegraphics[width=\columnwidth]{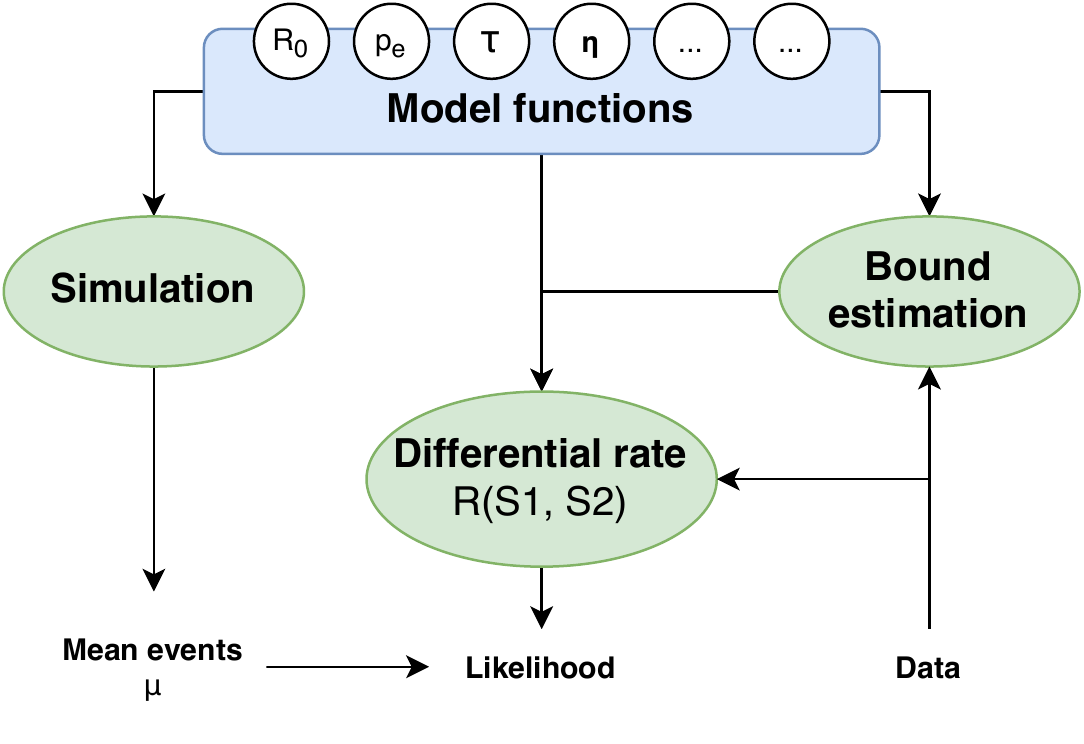}
    \caption{Sketch of \fd's functionality. Given data, \fd~estimates bounds for hidden variables such as the number of produced electrons. A TensorFlow representation of the data is used to compute the differential expected event rate for a signal or background source at the observed events. \fd~can also simulate events, which it uses to estimate the total number of expected events $\mu$ after efficiencies. These two ingredients make up the likelihood of equation \ref{eq:generic_logl}. All functionality uses a common set of model functions, such as electron lifetime $\tau$ or the input energy spectrum $\definable{R_0}(E)$. The user can customize these to make them depend on arbitrarily many observables, such as time or observed position. Most can also depend on particular unobserved variables, such as energy.
    }
    \label{fig:fd_source}
\end{figure}

Figure \ref{fig:fd_source} sketches the modeling functionality of \fd. Users can specify the desired physics through \emph{model functions} -- e.g.~how the electron lifetime and ER charge yield vary with space and time (and, for the latter, energy). Model functions will be highlighted in \definable{green} in the text below. Each can depend on an arbitrary number of observables such as $(x, y, z, t)$, and some can additionally depend on hidden variables, such as energy. The model functions are used for three main computations in \fd. 

First, given data (provided as a Pandas DataFrame \cite{pandas}), \fd~estimates reasonable ranges of the hidden variables that are summed over in the computation. A parameter \maxsigma~controls the width of the bounds: increasing it yields more accurate results for outlier events at the cost of computational speed. For this paper, we use $\maxsigma = 5$, to much the same effect as in the example in section \ref{section:outline}.

Second, \fd~uses the model functions inside the differential rate computation itself. We first convert the data to a series of TensorFlow tensors, segmented in batches of configurable size to control the memory required for the computation. Next, we compute a series of tensors that are multiplied together to yield the differential rate, as explained in section \ref{section:diffrate_outline} - \ref{section:final_signal_production}.

Finally, \fd~also contains an ordinary Monte Carlo code for simulating events using the same model functions. Besides its use in creating toy datasets, we will use it in section \ref{section:model_verification} to confirm that \fd's differential rate computation matches the result of an equivalent template-based method.

\begin{figure*}
    \centering
    \includegraphics[width=0.9\textwidth]{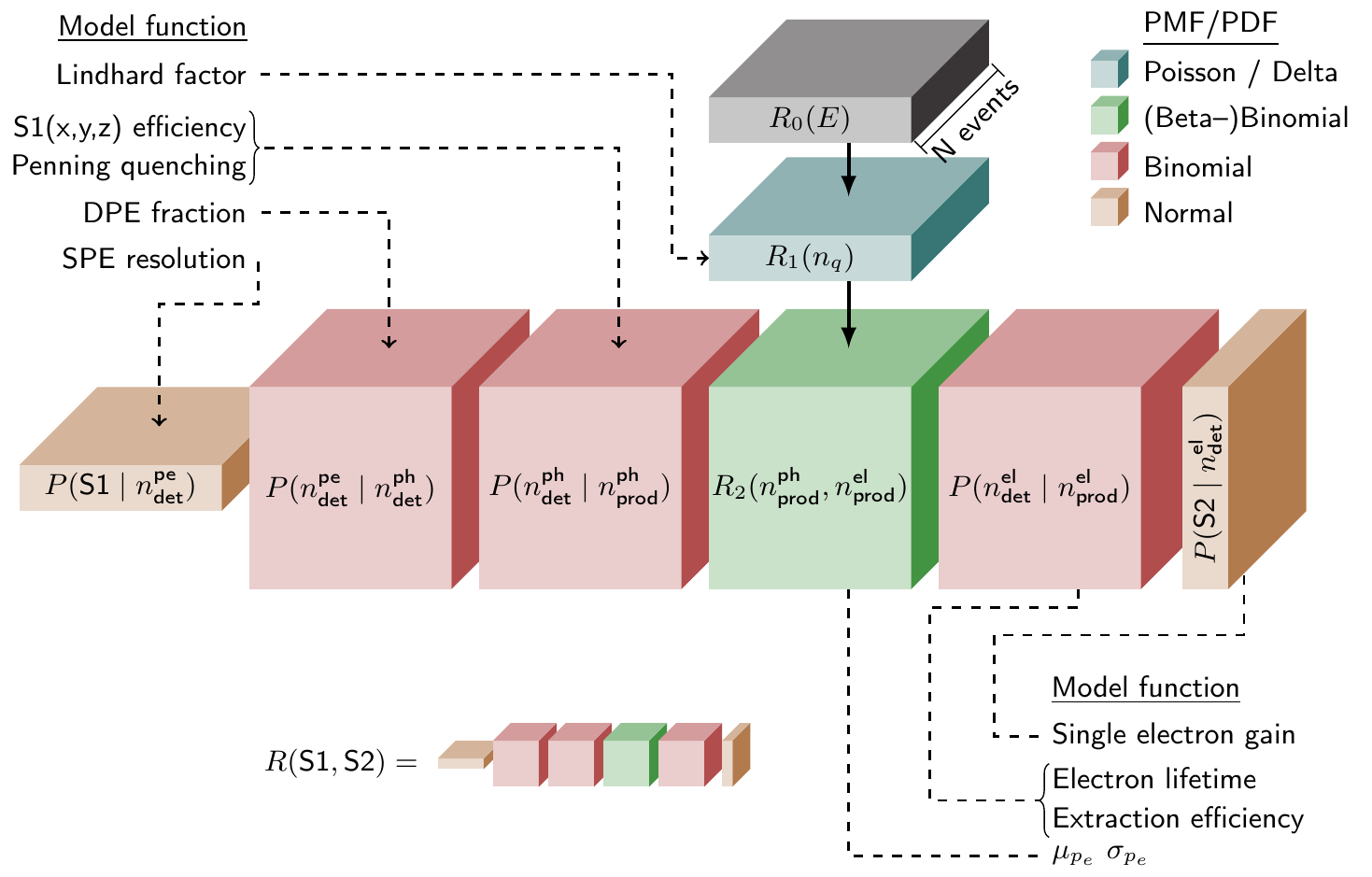}
    \caption{The LXe signal emission and TPC detector response model implemented in \fd. The differential expected event rate $R(\text{S1},\text{S2})$ at the observed event is computed as the matrix multiplication of the colored tensors on the lowest row. The figure also shows some of the model functions used in each step of the computation. The computation is `batched' over several events at once to improve performance, as indicated by the `N events' depth dimension. This is an implementation detail; the differential rate is computed independently for each event. The figure describes the computation of the differential rate due to a single source; it has to be repeated several times for models with multiple sources.}
    \label{fig:fd_diagram}
\end{figure*}

\subsection{Mean estimation}
\label{section:mu_estimation}

The simulation code also plays a role in \fd's inference. Besides differential rates, the likelihood in eq.~\ref{eq:generic_logl} uses the expected total number of events $\mu$. Computing this by integrating the differential rate computation over the space of possible observables would be very inefficient. Instead, we create a toy dataset with a configurable number of events (by default $10^5$) and count how many yield detectable signals that pass all efficiencies.

Like differential rates, the $\mu^j$ depend on shape parameters $\theta$: see eq.~\ref{eq:rate_multipliers}. \fd~is able to use the same method as classical template morphing: precompute $m^j$ at some particular $\theta$ values, then interpolate between the results while $\theta$ varies during the inference. This is not a significant computational burden, since a single number is much simpler to estimate and interpolate than a multi-dimensional template. For example, as discussed in \cite{morphing}, conventional (`vertical') interpolation performs poorly in modeling shifts in distributions, even in one dimension. For a single number such problems do not occur.

Interestingly, template-like interpolation of $m^j$ is not needed at all in many cases, or a very coarse estimate suffices. If the error on the estimate of $m^j(\theta)$ is well below the width of the external constraint on $r^j$, fits or limit constructions will converge to the correct values of the shape parameters $\hat{\theta}$, and the inaccuracy in $m^j(\theta)$ will be absorbed by the fitted rate multiplier $\hat{r}^j$ alone. After the fit, a single simulation can then be used to estimate $m^j(\hat{\theta})$ accurately, and we find the correct $\hat{r}^j$ by dividing the fitted $\hat{r}^j$ by the ratio of the accurate (post-hoc) and inaccurate (as used during the fit) estimates of $m^j(\hat{\theta})$.

Indeed, LXe TPC likelihoods generally have no or weak external constraints on the rate multipliers $r^j$ of the dark matter signal and the dominant internal ER background -- since the dark matter rate is the main parameter of interest, and the most sensitive measurement of the internal ER background usually comes from the data itself. For such a source,  pre-fit estimates of $m^j(\theta)$ are not needed at all. For some backgrounds, such as radiogenic neutrons, external constraints on $r^j$ are relevant, since their expected rate is too low to be effectively constrained by the data itself. However, for rare backgrounds, small inaccuracies in $m^j(\theta)$ will only have a marginal effect.

\subsection{Model structure}
\label{section:diffrate_outline}

In the remainder of this section, we will describe the ER/NR light- and charge emission and detector response model implemented in \fd. More precisely, we describe the structure in which several physics models can be implemented; the exact form of model functions such as light and charge yield can be chosen by collaborations themselves. \fd~also has an interface for users to change the core structure of the model, e.g.~to add extra factors or change assumed distributions fundamentally, which is described in our software documentation \cite{fd_github}. As mentioned above, we provide reasonable defaults inspired by NEST \cite{nest_paper, nest_software} and our previous work on XENON1T \cite{sr0_prl, analysis_paper_1, analysis_paper_2, aalbers_phd}. 

Throughout this section, an ``event'' will be taken to mean a single localized energy deposition in the LXe, causing a single S1 and a single S2 signal. At present \fd~only models single-scatter interactions, or multi-site scatters occurring sufficiently close together that they can be regarded as single scatters for all purposes. These comprise the dominant signals and backgrounds in LXe dark matter searches; resolvable multi-site scatters are efficiently rejected by data quality cuts \cite{analysis_paper_1, analysis_paper_2}.

Figure~\ref{fig:fd_diagram} illustrates the implementation of the differential rate calculation in \fd. In outline, we start with an input energy spectrum $\definable{R_0}(E)$. For each event, this is a vector whose elements range over different energies -- see section \ref{section:input_spectrum}. The computation is vectorized over a batch of events, indicated as the depth dimension in figure \ref{fig:fd_diagram}.
Since each event yields a stochastic number of detectable quanta $n_q$, we transform the expected spectrum at each event to a vector $R_1(n_q)$ over $n_q$ -- see section \ref{section:detectable_quanta}. Some quanta manifest as scintillation photons ($n^{\phot}_{\produ}$), others as drifting electrons ($n^{\elec}_{\produ}$) -- see section \ref{section:quanta_splitting}. Accounting for this, the differential rate becomes a matrix $R_2(n^{\phot}_{\produ}, n^{\elec}_{\produ})$. Some $n^{\phot}_{\dete}$ ($n^{\elec}_{\dete}$) of the produced photons (electrons) survive to be detected, as quantified by a matrix  $P(n^{\phot}_{\dete} | n^{\phot}_{\produ})$ ($P(n^{\elec}_{\dete} | n^{\elec}_{\produ})$) -- see section \ref{section:quanta_losses}. Finally, the detected photons (electrons) determine the S1 (S2) amplitude, as discussed in section \ref{section:final_signal_production}. In brief,  $P(\mathrm{S2} | n^{\elec}_{\dete})$ is assumed to be Gaussian, while for S1, two factors 
$P(n^{\phe}_{\det} | n^{\phot}_{\dete})$ and $P(\mathrm{S1} | n^{\phe}_{\dete})$ are involved, representing, respectively, the probability of detecting $n^{\phe}_{\dete}$ photoelectrons in the PMTs, then having these produce the measured S1 signal. Note that $P(\mathrm{S1} | n^{\phe}_{\dete})$ ($P(\mathrm{S2} | n^{\elec}_{\dete})$) is only a vector in $n^{\phe}_{\dete}$ ($n^{\elec}_{\dete})$ since the S1 (S2) value is already observed.
We multiply the $P$'s and $R_2$ together to get $R(\mathrm{S1}, \mathrm{S2})$, the differential expected event rate at the observed event.

\subsection{Input energy spectrum}
\label{section:input_spectrum}

\fd's starting point is an input energy spectrum, represented as a comb of delta functions. This is a natural input format for spectra specified as a finely (but possibly non-homogeneously) binned histogram.
Formally, flamedisx requires a differential rate $\definable{R_0}$, where
\begin{equation}\label{eq:input_difrate}
    \mu_0 = \rho \sum_E \int dx dy dz dt \definable{R_0}(E, x, y, z, t)
\end{equation}
is the total number of expected events in a hypothetical detector with perfect efficiency. Thus $\definable{R_0}$ has units of e.g.~$\mathrm{events}/(\mathrm{tonne} \mathrm{year})$. There is no $d\energysymbol$ in eq.~\ref{eq:input_difrate}; this is absorbed in $\definable{R_0}$.

For a homogeneous internal background, $\definable{R_0}$ is only a function of energy (and perhaps time); for dark matter, $\definable{R_0}$ is a function of energy and time (since dark matter signals are expected to have an annual modulation \cite{freese}); for external backgrounds, $\definable{R_0}$ can be a function of all observables.
Like all model functions, the $x, y, z, t$ have no particular special status, and the user can make $\definable{R_0}$ depend on an arbitrary number of observables.
Dependencies on hidden variables -- such as $E$ for $\definable{R_0}$ -- are fixed, since they determine the structure of \fd's computation.

\subsection{Detectable quanta}
\label{section:detectable_quanta}

The differential rate as a function of the number of quanta is:
\begin{equation}\label{eq:diffnq}
    R_1(n_{q}) = \sum_{\energysymbol} P(n_{q} \mid \energysymbol) \definable{R_0}(\energysymbol) .
\end{equation}
We omitted the dependence on $(x, y, z, t)$, and will do so below, since all model functions can depend on arbitrarily many observables.

The probability to generate $n_q$ quanta (scintillation photons and ionization electrons) given some deposited energy $\energysymbol$ is given by
\begin{align}\begin{split}
    \label{eq:pnq}
    P(n_{q} \mid \energysymbol)_{\text{ER}} &= \delta(n_{q} = \lfloor \energysymbol/ \definable{W} \rfloor) \\
    P(n_{q} \mid \energysymbol)_{\text{NR}} &= \text{Poisson}(n_q \mid \energysymbol \definable{\mathcal{L}}(\energysymbol) / \definable{W}),
\end{split}\end{align}
for ER and NR, respectively. Here $\delta$ is the Kronecker delta function, $\mathrm{Poisson}$ the Poisson probability mass function, $\mathcal{W} \approx \SI{13.8}{eV}$ the LXe work function, i.e. the energy required to generate one detectable ER quantum, and $\definable{\mathcal{L}}(\energysymbol)$ the mean fraction of NR energy used for making detectable quanta (rather than lost as heat). This is often called the Lindhard factor, though the user can specify any functional form of $\definable{\mathcal{L}}$, not just the one proposed by Lindhard \cite{lindhard_ancient}.

\subsection{Splitting over photons and electrons}
\label{section:quanta_splitting}

The differential rate of producing $n^{\phot}_{\produ}$ photons and $n^{\elec}_{\produ}$ electrons is given by:
\begin{multline}\label{eq:probprodu}
    R_2(n^{\phot}_{\produ}, n^{\elec}_{\produ}) = \sum_{n_{q}} P(n^{\elec}_{\produ} \mid n_{q}) R_1(n_{q}) \\
    \delta(n_{q}=n^{\phot}_{\produ} + n^{\elec}_{\produ}).
\end{multline}

The detected quanta will split binomially over photons or electrons. Thus $n^{\phot}_{\produ} = n_q - n^{\elec}_{\produ}$, and
\begin{equation*}
    P(n^{\elec}_{\produ} \mid n_{q}) = \text{Binom}(n^{\elec}_{\produ} \mid n_{q}, p_{e}),
\end{equation*}
where $\mathrm{Binom}$ is the binomial probability mass function and $p_e$ the fraction of detectable quanta that manifest as electrons. ERs show an overdispersion on top of this, commonly modeled as a Gaussian fluctuation on top of $p_{e}$, likely due variations in recombination \cite{lux_tritium, lux_yields, analysis_paper_2}. The Gaussian model cannot be physical, since it allows probabilities outside [0,1]. Instead, we assume the fluctuation is described by a beta distribution:
\begin{equation*}
    P(p_{e} \mid n_{q}) = \text{Beta}(n^{\elec}_{\produ} \mid \alpha, \beta) .
\end{equation*}
We can express the beta distribution's formal parameters $\alpha$ and $\beta$ in the mean $\definable{\mu_{p_{e}}}$ and standard deviation $\definable{\sigma_{p_{e}}}$ as
\begin{equation*}
    \alpha = \definable{\mu_{p_{e}}} \left( \frac{\definable{\mu_{p_{e}}}}{\definable{\sigma_{p_{e}}}^2} - \frac{\definable{\mu_{p_{e}}}^2}{\definable{\sigma_{p_{e}}}^2} - 1 \right),
    \quad
    \beta= \alpha (\frac{1}{\definable{\mu_{p_{e}}}} - 1)
\end{equation*}
Since the beta distribution is a conjugate distribution of the binomial, the process is equivalent to drawing $n^{\elec}_{\produ}$ directly from a beta-binomial distribution. Thus, our model for $n^{\elec}_{\produ}$ for ERs and NRs is:
\begin{align}\begin{split}
    P(n^{\elec}_{\produ} \mid n_{q})_{\text{ER}} 
        &= \text{BetaBinom}(n^{\elec}_{\produ} \mid n_{q}, \definable{\mu_{p_e}}(n_{q}), \definable{\sigma_{p_e}}(n_{q})) \\
    P(n^{\elec}_{\produ} \mid n_{q})_{\text{NR}} 
        &= \text{Binom}(n^{\elec}_{\produ}  \mid n_{q},  \definable{p_{e}}(n_{q})) .
    \label{eq:p_electron_combined}
\end{split}\end{align}
The parameters are functions of $n_{q}$ rather than $\energysymbol$. For ER, $\energysymbol$ and $n_q$ have an identity relation (see eq. \ref{eq:pnq}), so this is merely a change of notation. 
For low-energy NRs, it is an open question whether a direct dependence on $\energysymbol$ instead of $n_q$ would better fit the data. As low-energy NR tracks are much smaller than the electrons' thermalization length \cite{mozumder_yield, srim}, at the time of recombination, there should be no memory of $\energysymbol$ other than through the number of surviving excitons and electron/ion pairs $n_q$. Thus a direct dependency on $n_q$ may well be favored. On the other hand, post-thermalization recombination is not the only process involved.

Current models often apply a separate factor for suppression of the NR light yield due to Penning quenching \cite{lenardo_global_nr, mei_nr_2008}. At present, \fd~implements this by unphysically making the photon detection efficiency a function of the number of produced photons; a more elegant implementation would be to adjust the forms and parameters for $\definable{p_e}$ and the Lindhard factor $\definable{\mathcal{L}}$.


\subsection{Signal quanta losses}
\label{section:quanta_losses}

Some of the electrons and photons produce detectable signals in the PMTs, while others are lost. These efficiencies are usually position-dependent and include a variety of effects, such as absorption by impurities and lossy extraction through the liquid-gas interface for electrons, and imperfect collection and quantum efficiency for photons. Each signal photon and electron undergoes these processes individually, without interacting with other photons or electrons. Thus the probability of each photon to be detected (given a particular starting position, time, etc.) must be the same and independent of what happens to the other photon, and likewise for electrons. The fraction of detected photons (and likewise electrons) must therefore follow a binomial distribution.

Separately from the \emph{per-quanta} efficiencies $\definable{\eta^\phot}$ and $\definable{\eta^\elec}$, experiments have \emph{per-event} detection efficiencies. For example, two-photon S1s are often excluded from analysis, since these are likely to arise from accidental coincidence of PMT dark counts. These efficiencies are usually parametrized as functions of either the raw S1 or S2 signal amplitude (see the next section) or the number of detected photons and electrons. We call the latter $\definable{\zeta^\phot}$ and $\definable{\zeta^\elec}$, respectively.

Given $n^{\phot}_{\produ}$ produced photons ($n^{\elec}_{\produ}$ electrons), the probability of detecting $n^{\phot}_{\det}$ photons ($n^{\elec}_{\dete}$ electrons) is thus:
\begin{align}\begin{split}
    \label{eq:p_detection}
    P(n^{\phot}_{\dete} \mid n^{\phot}_{\produ}) 
        &= \definable{\zeta^{\phot}}(n^{\phot}_{\dete}) \text{Binom}(n^{\phot}_{\dete} \mid n^{\phot}_{\produ}, \definable{\eta^\phot}) \\
    P(n^{\elec}_{\dete} \mid n^{\elec}_{\produ})
        &= \definable{\zeta^{\elec}}(n^{\elec}_{\dete}) \text{Binom}(n^{\elec}_{\dete} \mid n^{\elec}_{\produ}, \definable{\eta^\elec}) .
\end{split}\end{align}

\subsection{Final signal production} 
\label{section:final_signal_production}
For sufficiently small S1s, the non-Gaussian response of PMTs to single photons is relevant. PMTs in LXe TPCs have a probability
$\definable{p_\mathrm{DPE}} \approx 0.2$
of producing two rather than one photoelectron for each detected LXe scintillation photon \cite{faham_dpe}. \fd~accounts for this with a binomial factor:
\begin{equation}
    P(n^{\phe}_{\dete} \mid n^{\phot}_{\dete}) = \text{Binom}(
    n^{\phe}_{\dete} - n^{\phot}_{\dete} |  n^{\phot}_{\dete}, \definable{p_\mathrm{DPE}} ) ,
\end{equation}
where $n^{\phe}_{\dete}$ is the number of detected photoelectrons in response to $n^{\phot}_{\dete}$ detected photons. The factor is set to zero for $n^{\phe}_{\dete} < n^{\phot}_{\dete}$. This binomial cannot be combined with the one in eq.~\ref{eq:p_detection}, because it is evaluated at $n^{\phe}_{\dete} - n^{\phot}_{\dete}$, not $n^{\phe}_{\dete}$. 

We only apply this factor for S1, because S2s signals are large enough ($> 100$ photoelectrons) that the non-Gaussian nature of the double-photoelectron emission effect is irrelevant. The double-photoelectron effect affects the interpretation of the Gaussian S2 response parameters (described below), but not the functional form of the response.

The observed S1 and S2 signal size given a number of detected quanta -- photoelectrons for S1, and extracted electrons for S2 -- is assumed to follow a Gaussian distribution. We thus have:
\begin{align}\begin{split}
    P(\mathrm{S1} \mid n^{\phe}_{\dete})
        &= \definable{\xi_\mathrm{S1}} \text{Normal}(\mathrm{S1} \mid \mu_\mathrm{S1}(n_\dete^\phe) ,
        \sigma_\mathrm{S1}(n_\dete^\phe)) \\
    P(\mathrm{S2} \mid n^{\elec}_{\dete})
        &= \definable{\xi_\mathrm{S2}} \text{Normal}(\mathrm{S2} \mid \mu_\mathrm{S2}(n^{\elec}_{\dete}),  \sigma_\mathrm{S2}( n^{\elec}_{\dete}) ),
    \label{eq:sproduction}
\end{split}\end{align}
with $\mu_\mathrm{S1}$ and $\sigma_\mathrm{S1}$ the mean and standard deviation S1 area in PE produced by a single photoelectron, $\definable{\xi^\mathrm{S1}}$ the per-event detection efficiencies parametrized as function of S1 (and thus not included in $\definable{\zeta^\phot}$), and similar quantities for electrons and S2s.

Assuming a linear detector response to photoelectrons and extracted electrons, the final signal means and standard deviations are fully specified in terms of those of a single quantum:
\begin{align}\begin{split}
\mu_\mathrm{S1}(n_\dete^\phe) &= \definable{\mu_{1}^\mathrm{S1}} n_\dete^\phe \\
\sigma_\mathrm{S1}(n_\dete^\phe) &= \definable{\sigma_1^\mathrm{S1}} \sqrt{n_\dete^\phe} \\
\mu_\mathrm{S2}(n_\dete^\elec) &= \definable{\mu_{1}^\mathrm{S2}} n_\dete^\elec \\
\sigma_\mathrm{S2}(n_\dete^\elec) &= \definable{\sigma_1^\mathrm{S2}} \sqrt{n_\dete^\elec}
\end{split}\end{align}
with $\mu_1^{S1}$ and $\sigma_1^\phe$ the mean and standard deviation of the S1 area produced per single detected photoelectron, respectively, and similar quantities for S2.

\clearpage

\section{Results}
\label{section:results}

\begin{figure}
    \centering
    \includegraphics[width=\columnwidth]{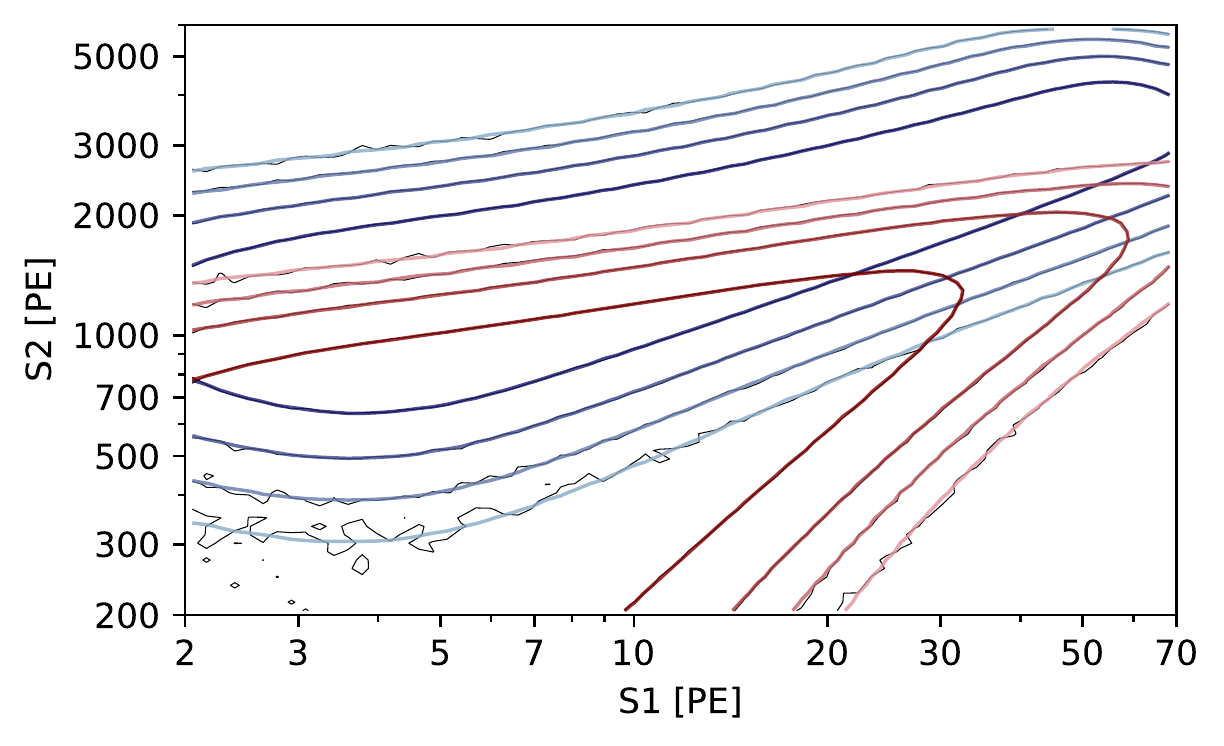}
    \caption{Contours in (S1, S2) enclosing, from bright to faint colors, 90\%, 99\%, 99.9\%, and 99.99\% of events with the highest differential rates, for a $0-\SI{10}{keV}$ flat ER source (blue) and a \SI{30}{GeV/c^2} WIMP (red), at a single event position and time. The colored contours show the result from \fd; thin black lines show the result from a template with $10^8$ simulated events, described further in the text.}
    \label{fig:model_verification}
\end{figure}

\subsection{Model verification}
\label{section:model_verification}

A first and essential check is to verify that \fd's direct differential rate computation (figure \ref{fig:fd_diagram}) produces the same result as a lookup in finely-binned template created from a high-statistic simulation. A divergence would point to an error in \fd. 

Computing even a single 6-dimensional $(\tsone, \tstwo, x, y, z, t)$ histogram with sufficient statistics for this check would be a significant undertaking. It is much more practical to fix the simulated event positions and time,
so only a two-dimensional (S1, S2) histogram is needed for comparison. We take the (S1, S2) histogram to have 70 bins in both S1 and S2, uniformly increasing in $\log S1$ and $\log S2$ respectively to better capture the quickly-varying region at low energy. We fill the histogram with $10^8$ simulated events, in accordance with (or excess over) current practice.

Figure \ref{fig:model_verification} compares the \fd~model against the high-statistic template, for two sources: a $0-\SI{10}{keV}$ flat ER background and a \SI{30}{GeV/c^2} WIMP, positioned at the center of a XENON1T-like detector. We see that \fd~agrees with the template deep into the tails of the distribution, until the high-statistics template becomes unreliable due to statistical errors. The result of the two methods should be exactly the same only in the limit of an infinitely large simulation, a histogram with infinitesimally small bins, and $\maxsigma \rightarrow \infty$.

It is simple to repeat this test, though it should not be necessary often. Users will frequently change
only the model functions, which are shared between the simulator and the direct computation. Changes in these cannot by themselves cause a divergence if the model is implemented correctly.

\subsection{Single-event discrimination}

\fd~can boost the physics potential of LXe detectors by making a full, undiscretized $(\mathrm{S1}, \mathrm{S2}, x, y, z, t)$ likelihood viable, in a robust analysis with several correlated nuisance parameters. In this section, we will show the benefits of this, by comparing \fd's likelihood with a two-dimensional $(\mathrm{cS1}, \mathrm{cS2})$ likelihood.

A fully specified model is necessary for this comparison. 
We used the ER and NR models described in \cite{analysis_paper_2} (reparametrized to fit the \fd~structure) and the S1 and S2 efficiency maps from \cite{pax}. The exact model functions can be inspected along with the the \fd~source code at \cite{fd_github}. 
We assume a fiducial volume of \SI{5}{tonnes}, an electron lifetime of \SI{500}{us}, a TPC length of \SI{1.5}{m}, a homogeneous drift field and perfectly stable detector conditions.

\begin{figure}
    \centering
    \includegraphics[width=\columnwidth]{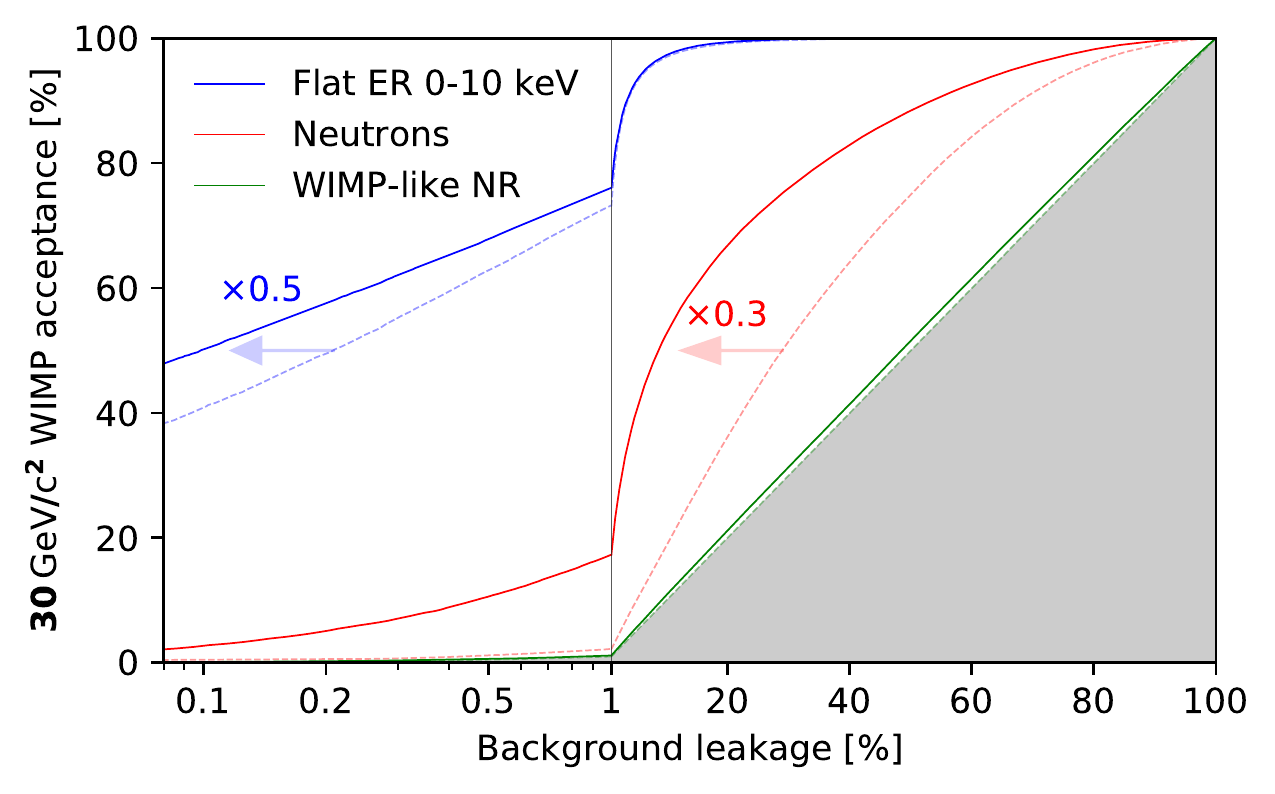}
    \includegraphics[width=\columnwidth]{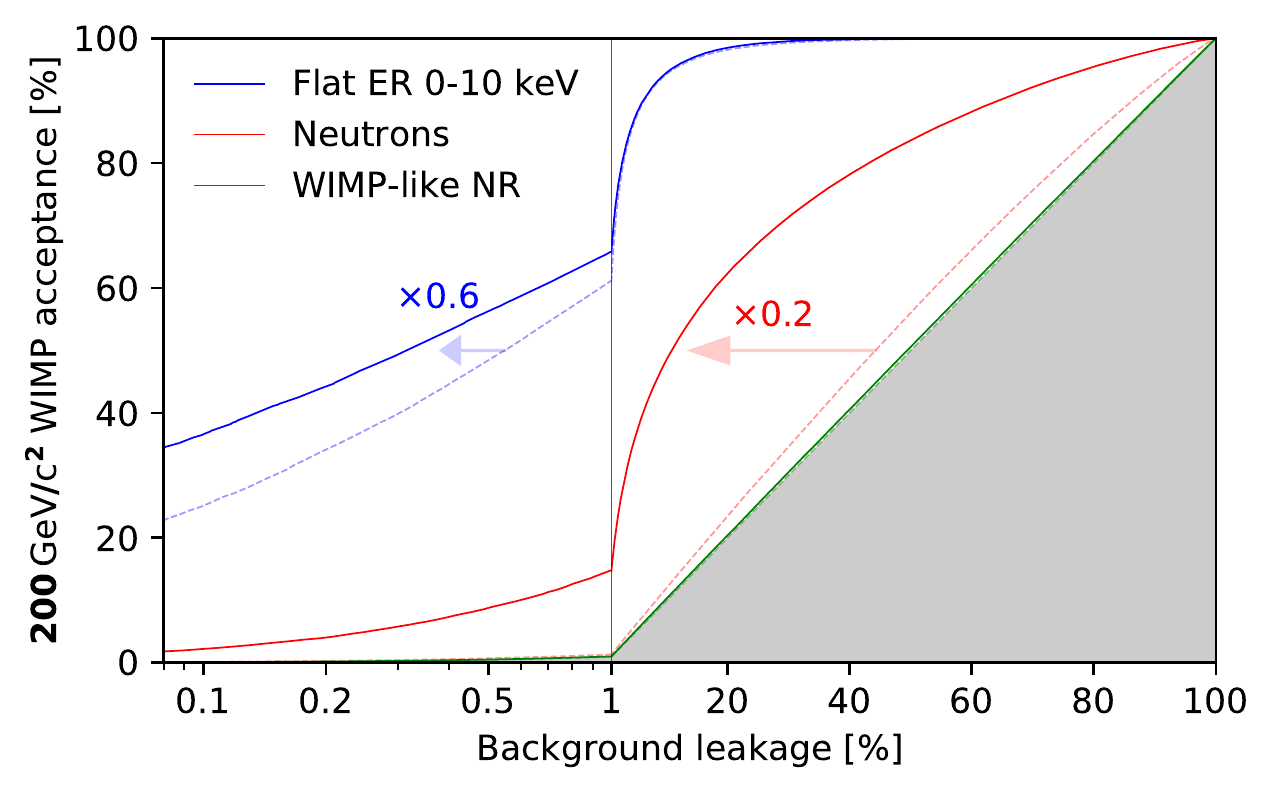}
    \caption{Discrimination of different types of backgrounds from a \SI{30}{GeV/c^2} (top panel) and \SI{200}{GeV/c^2} (bottom panel) WIMP signal in \fd~(solid) and a two-dimensional template-based likelihood (dashed), using exactly the same emission, detector response, and analysis efficiency assumptions. Blue: flat-spectrum homogeneous ER background from $0-\SI{10}{keV}$; Red: radiogenic neutron background with a spatially varying rate; Green: hypothetical homogeneous NR background with the same energy spectrum as the WIMP signal, but a time-averaged rather than an annually modulating spectrum. Note the switch from a linear to a logarithmic scale below $1\%$ background leakage. Arrows show the background reduction at $50\%$ signal acceptance.}
    \label{fig:roc_curves}
\end{figure}

Using this model, we compute two-dimensional $(\mathrm{cS1}, \mathrm{cS2})$ ER and NR templates, using the same binning discussed above, except that it is linear in $\mathrm{cS1}$ to match existing practice. We fill each template with $10^8$ events each, then use the templates as well as \fd's direct computation to estimate the differential rate for simulated events. The ratio of expected differential rates of two sources is the Neyman-Pearson optimal discrimination test statistic to discriminate single events.

Figure \ref{fig:roc_curves} shows several discrimination efficiencies (``ROC plots'') using this statistic. Specifically, it shows, for different backgrounds, the fraction of remaining expected WIMP signal events if a given fraction of the background must be removed. We see that, at 50\% signal acceptance, \fd~effectively reduces the dominant homogeneous flat ER background by a factor two. This is almost entirely due to separating different $z$ regions of the detector, which have different S2 resolution (and thus different discrimination power) due to the finite electron lifetime. The neutron background is discriminated by an even larger factor, due to its spatial dependence within the assumed fiducial volume. The WIMP signal's annual modulation contributes almost nothing to this discrimination: a hypothetical homogeneous NR background with the same mean energy spectrum as the signal (but without modulation) can barely be distinguished from the signal.

We stress that the same results would be obtained in a traditional template-based method with densely binned high-dimensional templates, if it would be feasible.

\subsection{Speed}
\label{section:speed}

\begin{figure}
    \centering
    \includegraphics[width=\columnwidth]{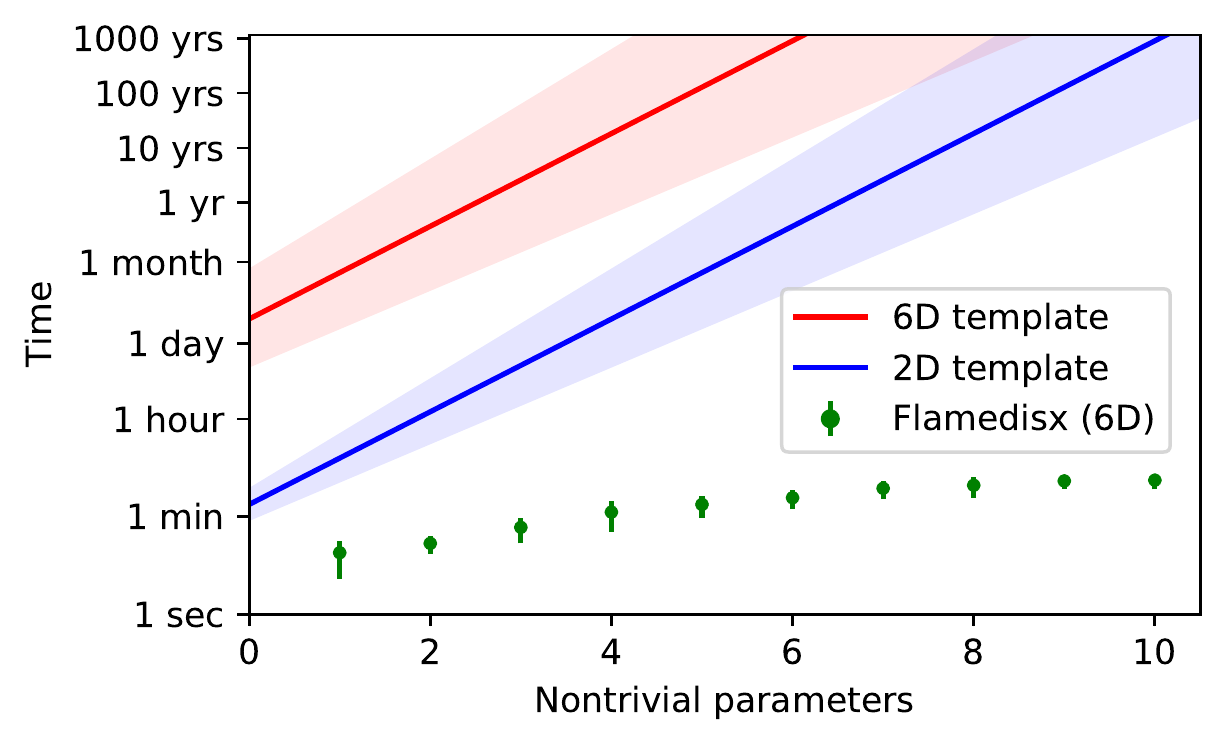}
    \caption{Lines: estimated time required to compute templates for a model as a function of the number of parameters that affect the observable distribution in a potentially correlated manner. Blue shows a two-dimensional $(\mathrm{cS1}, \mathrm{cS2})$ model, red a 6-dimensional $(\mathrm{S1}, \mathrm{S2}, x, y, z, t)$ model. The bands bracket optimistic and pessimistic assumptions detailed in the text. Green dots: time required to fit a 1000-event ER calibration dataset in \fd, as a function of the polynomial order of the ER electron yield model. The error bars are statistical errors from repeating the fit on different toy datasets.
    }
    \label{fig:benchmarks}
\end{figure}

\fd~has a fundamentally different performance profile than classic template-based likelihoods. For templates, filling the histograms with simulated events is usually the most expensive step: the number of events needed grows exponentially with the histogram dimensionality and the number of potentially correlated parameters that shape the distributions. \fd~does not use templates, and thus requires essentially no precomputation -- only $\mathcal{O}(\SI{10}{s})$ of TensorFlow graph construction, which does \emph{not} need to be repeated when a new (toy) dataset is considered.

Figure \ref{fig:benchmarks} shows an estimate of the required template computation time for a two-dimensional ($\mathrm{cS1}, \mathrm{cS2}$) and a six-dimensional ($\mathrm{S1}, \mathrm{S2}, x, y, z, t$) likelihood. For the central model (bands) we assumed a simulator capable of producing an accurate 2D-template model in 100 (50-200) seconds
\footnote{To see cS1/cS2 discrimination at the one in a thousand level, we need $\mathcal{O}(10^6)$ events per cS1 bin. Having $\mathcal{O}(100)$ cS1 bins thus means we need $\mathcal{O}(10^8)$ events to fill a template accurately. Assuming the simulator can create $\mathcal{O}(10^6)$ events per second -- flamedisx's built-in simulator achieves about half this speed on currently common CPUs -- implies that $\mathcal{O}(100)$ seconds are needed to construct a 2-dimensional template.}
and 7 (5-10) templates per parameter. For the 6D-template model, we assume 7 (5-10) bins each for ($x, y, z$ and $t$). Clearly, with more than a handful of nontrivial parameters, even two-dimensional templates become unwieldy, and six-dimensional template grids require computation times measured on geological scales.

After precomputation, template-based likelihoods are relatively fast: computing differential rates only requires a few lookups and simple interpolation. \fd~must instead do a new differential rate computation for each event, which is generally the rate-limiting step.

On a single Tesla K80 GPU, \fd's differential rate computation runs at $\roughly \SI{3}{events/ms}$ for events from a $0-\SI{10}{keV}$ ER model. NR models are nearly three times faster to evaluate, mainly because they lack the beta-binomial in equation \ref{eq:p_electron_combined}. Thus, the likelihood of a $\mathcal{O}(\SI{1000}{event})$ dataset under a model with both ER and NR sources can be computed in about half a second. \fd's differential rate computation can also run on a CPU, but then it is $\mathcal{O}(10^2)$ times slower.

During inference, the likelihood must be computed at many different points in parameter space. \fd~ exploits TensorFlow's automatic differentiation to compute the gradient and (optionally) the Hessian in parameter space. With this information, high-dimensional optimization needs far fewer iterations, and becomes more robust. Figure \ref{fig:benchmarks} shows the mean duration of fitting an ER model to a 1000-event ER calibration dataset. Specifically, we fitted polynomial models of different orders for $\mu_{p_e}(n_q)$ -- effectively controlling the mean ER charge yield as a function of energy.

Figure \ref{fig:benchmarks} is clearly not a level comparison of \fd~and a template-based inference framework: it shows an estimated precomputation time on a CPU for the template method, and a per-fit measured time on a GPU for \fd. We juxtapose the two pieces of information to highlight the fundamentally different scaling with the number of nontrivial parameters. A template-based framework could be ported to GPUs, but the only benefit would be to allow, perhaps, two or three more dimensions or parameters within the same computation time frame.

\subsection{Physics reach improvement}

\begin{figure}[t]
    \centering
    \includegraphics[width=\columnwidth]{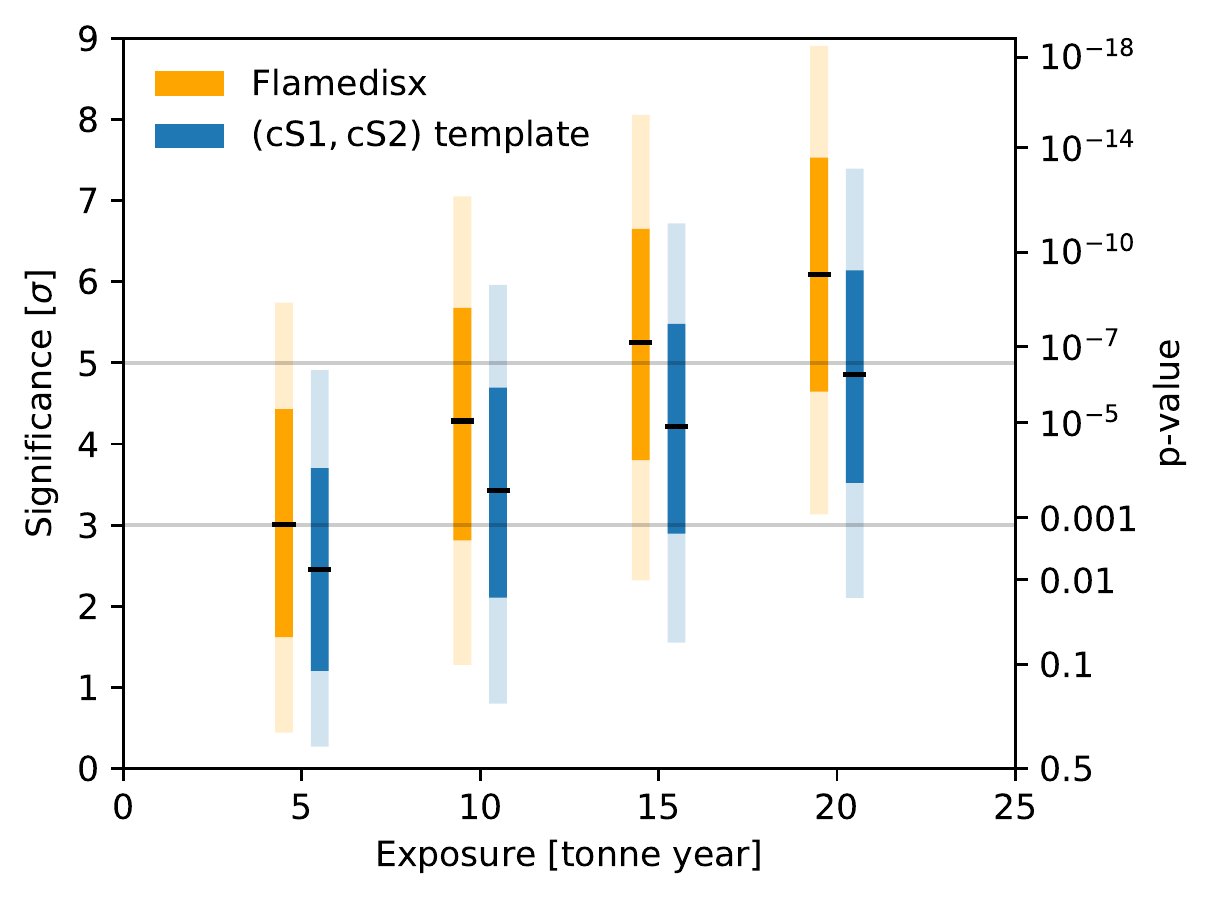}
    \caption{Median (black lines) and $1$ and $2\sigma$ bands of the asymptotic discovery significance for a $\sigma = \SI{2e-47}{cm^2}$, \SI{200}{GeV/c^2} WIMP ($\roughly \SI{1.4}{events/(tonne \times year)}$ expected) for different exposures of a XENONnT/LZ-like example experiment, using a traditional (cS1, cS2) likelihood (blue) and \fd's differential rate computation on the same \num{1e5} toy datasets. Significances are calculated at 5, 10, 15, and 20 tonne year; slight horizontal offsets are for visualization.
    }
    \label{fig:sensitivity}
\end{figure}

The improved background discrimination translates into an improved physics reach. 

One way to measure the physics reach is by the expected discovery significance, i.e.~the p-value of the background-only hypothesis, for one particular assumed WIMP signal.
Discovery significances can be computed using the profile likelihood test statistic:
\begin{equation}
    t_0 = -2 \log \frac{L(\sigma = 0, \hat{\hat{\theta}})}{L(\hat{\sigma}, \hat{\theta})} ,
\end{equation}
where $\sigma$ is the WIMP cross section, $\theta$ represents nuisance parameters (here only the rate of the internal ER background), $\hat{\sigma}, \hat{\theta}$ denote the global best-fit value, and $\hat{\hat{\theta}}$ the best fit of $\theta$ conditional on $\sigma = 0$. For sufficiently large exposures, as well as certain other technical conditions \cite{beyond_wilks}, $t_0$ is asymptotically distributed as $\frac{1}{2} \delta(0) + \chi^2_{\nu=1}$, with $\delta$ the Dirac delta function and $\chi^2_{\nu=1}$ the chi-squared distribution with one degree of freedom.
We verified that the distribution of $t_0$ is well-described by this approximation for \SI{200}{GeV/c^2} WIMPs and exposures of \SI{10}{tonne year} and higher, using fits to background-only toy MCs. Specifically, a $3 \sigma$ asymptotic significance corresponds to a $3.0 \pm 0.2 \sigma$ true significance, for both \fd~and the $(\mathrm{cS1}, \mathrm{cS2})$ template method.

Figure \ref{fig:sensitivity} shows the expected asymptotic discovery significances for a $\sigma = \SI{2e-47}{cm^2}$, \SI{200}{GeV/c^2} WIMP in different exposures of the XENONnT/LZ like test model discussed above. This model is not excluded by current experiments \cite{sr1_prl}, but well within range of the next generation of detectors \cite{lz_sensitivity, nt}. We assumed a flat ER background level of $\SI{75}{events/(tonne \times year)}$ and a radiogenic NR background of $\roughly\! \SI{0.04}{events/(tonne \times year)}$.
Clearly, \fd~gives a substantial advantage over using a classic $(cS1, cS2)$ likelihood. With a $\roughly \SI{5}{tonne}$ fiducial volume, it cuts the time needed to get a $5 \sigma$ discovery for this model by around a year.
More details on this projection can be found in \cite{pelssers_phd}.

We stress that our model should be regarded as a test model only. Accurate projections of the performance of future detectors are more complex and rely on information available only to collaborations themselves. For example, we omitted the accidental coincidence and coherent neutrino nucleus-scattering background completely, and our neutron background spectral shape and its spatial rate dependence are chosen to roughly approximate published figures \cite{1t_mc, sr1_prl}.

The physics reach can also be characterized by the median exclusion sensitivity in case dark matter does not exist. In this case, background discrimination is much less relevant, since the dark matter models probed are by definition at the statistical limit of what the experiment is capable of distinguishing. Only in the extreme background-limited case, where the sensitivity scales with the square root of exposure, do discrimination improvements (and \fd) yield a proportional improvement of the exclusion sensitivity. LXe DM searches so far have usually stopped running to build a bigger detector well before reaching this limit. 

\section{Discussion}

\subsection{Validity of the model}

As discussed in section \ref{section:method}, our model structure is not exactly equivalent to models previously used in the field, but designed to have ample expressive power to fit real data. Since LXe collaborations have not released their raw data to the public, we cannot report on our efforts to verify this here. For future \fd~versions, we hope to implement defaults that more closely track NEST \cite{nest_software} for a variety of detector conditions.

We do \emph{not} suggest experiments use the  defaults of \fd~out of the box, but that they use \fd~to fit their calibration data first. Regardless of which modeling framework is used, assessing its goodness of fit is critical. \fd's integrated simulator can be used to produce lower-dimensional templates for use in non-parametric tests, e.g. chi-squared or Kolmogorov-Smirnov. If the fit is poor, users can re-fit to calibration data, change the model functions, or even change the model structure fundamentally.

A standard way to address (potential) mismodeling is to add nuisance parameters for uncertain aspects of the model. \fd~helps here, because it can handle more correlated nontrivial uncertainties than a MC-template based framework. Finally, \fd~is a single framework capable of both calibration data fitting and the final scientific inference. Calibration data can also be fitted simultaneously with the final inference, eliminating the sometimes difficult choice of which uncertainties to propagate, and thereby another source of potential mismodeling.

Using a `six-dimensional likelihood' might give some readers pause. Would this mean that orders of magnitude more calibration data must be collected to verify that the model fits? This is not the case, for several reasons.
First, we must not confuse the number of observables with the number of parameters in the model.
\fd, like an MC simulation, can use as many or few parameters in its functions as the user wants, which determines whether it will under- or overfit the data. \fd~looks at more observables (six) than a two-dimensional template for the same number of events, so it is a more sensitive instrument to decide between models. If anything, using \fd~would mean that \emph{less} calibration data is needed by an experiment.

Moreover, a two-dimensional template is equivalent to a six-dimensional template that assumes the response is constant over four of the dimensions. A model that actually looks at six observables only has to improve on this rather low bar to be superior -- e.g. by accounting for simple and easily verified effects, such as the reduction in S2 resolution by electron lifetime. Finally, since \fd~obviates computationally expensive template morphing, it becomes simpler to add additional nuisance parameters representing model uncertainties to the inference. Properly used, this will increase the robustness of results.

\subsection{Impact on real experiments}
\fd's advantage in physics reach is entirely due to allowing a full undiscretized ($\mathrm{S1}$, $\mathrm{S2}$, x, y, z, t) likelihood. We compared \fd~against a $(\mathrm{cS1}, \mathrm{cS2})$ likelihood above (e.g.~figure \ref{fig:sensitivity}), but LXe dark matter search likelihoods often incorporate space and time dimensions already in a limited way. For two reasons, we still believe using \fd~would significantly improve the physics reach of the imminent generation of experiments.

First, the space and time dimensions are very sparsely discretized in current likelihoods. That is, the histogram templates may cover more than two dimensions, but the number of bins in the spatio-temporal dimensions is low.
The likelihood itself is still unbinned, but less accurate than it would be with an undiscretized model.
XENON1T initially used a 2-d $(\mathrm{cS1}, \mathrm{cS2})$ likelihood \cite{sr0_prl}. In XENON1T's full science run, a sparse spatial discretization was used, mostly in the $r$ dimension \cite{sr1_prl}. LUX's likelihood was discretized in four $z$ segments and four time segments \cite{lux_complete}. PandaX-II used 18 time segments \cite{pandax_54}. LZ's sensitivity projection assumes a 2-d $(\mathrm{cS1}, \mathrm{cS2})$ likelihood \cite{lz_sensitivity}.

Second, we assumed constant detector conditions and homogeneous drift and extraction fields in our example experiment. Neither of these were realized in the last generation of detectors \cite{sr1_prl, lux_complete}. For detectors with variable conditions, space and time dimensions in the likelihood are more important, so the benefit of using \fd~is greater.

\section{Conclusions and outlook}
\label{section:outlook}

We described a new framework for computing LXe TPC likelihoods, \fd, which replaces simulation-based templates with a direct differential rate computation implemented in TensorFlow. This enables high-dimensional undiscretized likelihoods, which will increase the physics potential of LXe detectors. It also enables consideration of more correlated nuisance parameters, leading to more robust results.

We hope \fd~will be useful for experimental collaborations. 
The \fd~source code is released at \cite{fd_github} under a permissive open-source license, and will continue to be developed.
As members of XENON, we are particularly looking forward to possible application of \fd~in the upcoming XENONnT experiment \cite{nt}. Eventually, we hope the methods used in \fd~might inspire similar changes in likelihoods used by other types of particle physics experiments.

\subsection{Acknowledgements}
The authors would like to thank Knut D. Morå, Fei Gao, and others in the XENON Collaboration for useful discussions. The authors gratefully acknowledge support from the Knut and Alice Wallenberg Foundation and the Swedish Research Council.

\bibliographystyle{apsrev4-1}
\bibliography{references}



\end{document}